\documentstyle[12pt]{article}
\def\preprint#1{\thispagestyle{empty}~\newline\vspace*{-22.65mm}
\begin{flushright}\begin{tabular}{l} #1 \end{tabular}
\end{flushright}\vspace{1cm}}
\newlength{\inda}
\newlength{\indb}

\def\euc{{\sc e}}
\def\ym{{\sc ym}}

\begin{document}

\preprint{
\begin{tabular}{l}
gr-qc/9805010\\ Class. Quantum Grav. 15 (1998) 3763
\end{tabular}}
\markright{Thiemann transform for gravity with matter fields}

\begin{center}
{\large \bf Thiemann transform for gravity with matter fields}
\\[5mm] {Luis J Garay and Guillermo A Mena Marug\'{a}n}
\\[3mm] {\it  Instituto de Matem\'{a}ticas y F\'{\i}sica Fundamental, CSIC,
\\ Serrano 121, 28006 Madrid, Spain}
\\[3mm] 10 September 1998
\end{center}

\begin{abstract}

The generalized Wick transform discovered by Thiemann provides a
well-estab\-lished relation between the Euclidean and Lorentzian
theories of general relativity. We extend this Thiemann
transform to the Ashtekar formulation for gravity coupled with
spin-1/2 fermions, a non-Abelian Yang-Mills field and a scalar
field. It is proved that, on functions of the gravitational and
matter phase space variables, the Thiemann transform is
equivalent to the composition of an inverse Wick rotation and a
constant complex scale transformation of all fields. This result
also holds for functions that depend on the shift vector, the
lapse function and the Lagrange multipliers of the Yang-Mills
and gravitational Gauss constraints, provided that the Wick
rotation is implemented by means of an analytic continuation of
the lapse. In this way, the Thiemann transform is furnished with
a geometric interpretation. Finally, we confirm the expectation
that the generator of the Thiemann transform can be determined
just from the spin of the fields and give a simple explanation
for this fact.
\end{abstract}

\section{Introduction}

A well-established relation between the Lorentzian and Euclidean
theories of vacuum general relativity was found recently by
Thiemann \cite{th96}. This relation is provided by a generalized
Wick transform (which, from now on, we will call the Thiemann
transform), defined as an automorphism on the algebra of
functions on the gravitational phase space that preserves
Poisson brackets. The Thiemann transform generally maps real
functions to complex ones, so that it modifies the reality
conditions. An extension of this transform to the case when
certain types of matter sources (scalar and electromagnetic
fields) are present was proposed by Ashtekar \cite{as96}.

The Thiemann transform was originally formulated in the Ashtekar
formalism for general relativity, although it can also be
employed, for example, in geometrodynamics and in the triad
formalism \cite{as96,me98}. In Ashtekar's approach to Lorentzian
quantum gravity, it seems necessary to impose complicated
reality conditions because the Ashtekar connection is complex
for Lorentz\-ian metrics \cite{as91,g3}. This problem can be
overcome using the Thiemann transform, which maps the Lorentzian
to the Euclidean theory of gravity. In fact, the reality
conditions are easy to impose in the Euclidean sector since the
Ashtekar variables can be chosen to be real in this
case\footnote{This point is discussed in section 5, including
the case when fermions are present.}. Furthermore, since the
Thiemann transform maps the Euclidean constraint functionals to
the Lorentzian ones, it turns out that this transform allows to
obtain the Lorentzian quantum states starting from the Euclidean
quantum theory \cite{th96,as96}. In this sense, the transform
discovered by Thiemann enables one to work with real Ashtekar
variables when quantizing general relativity.

The role of the Thiemann transform in quantum gravity is similar to
that played by the Wick rotation in quantum field theory in flat
spacetime. Both transformations provide a map between a theory that
is mathematically difficult to handle and another theory that is, in
principle, more manageable. For instance, the Lorentzian path
integral in quantum field theory diverges because of the oscillatory
character of the integrand $e^{iS}$. The Wick rotation changes this
integrand into $e^{-I}$, where $I$ is the Euclidean action. Then,
the resulting Euclidean path integral turns out to converge for most
matter fields, namely, scalar, Yang-Mills and fermionic fields. In
general relativity, the Euclidean path integral is still ill-defined
because the gravitational action is unbounded \cite{7,9}.
Nevertheless, it has been proposed that this integral can also be
made convergent (at least in the one-loop approximation) by
integrating the conformal factor over an appropriate complex contour
\cite{9,10}. Similarly, the problem of dealing with the complicated
reality conditions of Lorentzian gravity in the Ashtekar formalism
is circumvented with the help of the Thiemann transform. In this
way, one arrives at a theory in which Ashtekar variables can be
treated as real and that, remarkably, happens to describe Euclidean
general relativity.

Despite this analogy between the Thiemann transform and the Wick
rotation, it was initially far from clear that a stronger
relation between them could actually exist \cite{as96}. Indeed,
while the Wick rotation can be regarded as an analytic
continuation that, from a geometric point of view, is based on a
complexification of time, the Thiemann transform, defined as a
map on the algebra of functions on phase space, seems to lack a
proper geometric interpretation. Such an interpretation would be
a helpful guideline for the application of the Thiemann
transform to the study of problems of interest in gravity and
cosmology.

In order to clarify this point, one of the authors showed
recently \cite{me98} that the effects of the Thiemann transform
on the line element, constraints, and action for vacuum general
relativity could be regarded as coming from an inverse Wick
rotation and a constant complex scale transformation of the
gravitational field. This rescaling amounts to a constant
complex conformal transformation of the spacetime metric. As a
consequence, it is actually possible to attain a geometric
interpretation for the Thiemann transform. Furthermore, it was
also proved that the equivalence between the action of the
Thiemann transform and the composition of an inverse Wick
rotation with a complex conformal transformation applies as well
to any function on phase space and to the lapse and shift
functions, provided that the Wick rotation is implemented by
analytically continuing the lapse from positive real to negative
imaginary values\footnote{This rotation can be viewed as a
continuation in a signature parameter. Some papers that consider
this analytic continuation of the signature are are given in
\cite{g9}.} \cite{g8,hd}. The main aim of the present paper is
to show that this geometric interpretation continues to be valid
in the presence of matter sources. In this case, the constant
conformal transformation that is involved in the Thiemann
transform becomes a constant complex scale transformation of all
the fields that is uniquely determined by their respective
spins.

Let us briefly review the Hamiltonian formulation for general
relativity with matter fields in the Ashtekar formalism. The
gravitational phase space can be described by the fields $(a_a,
\sigma^{a})$, defined on a 3-manifold. Here,
$a_{aA}^{\hspace{\inda}B}$ is the Ashtekar connection and
$\sigma_{\;A}^{a\;\;B}$ are the densitized $SU(2)$ soldering
forms \cite{as91}. As matter content, we will consider a massive
scalar field $\phi$, massive spin-$1/2$ fields $\xi_A$and $\bar
\eta_A$, and a Yang-Mills connection $A_a$, where the internal
Yang-Mills indices have been suppressed. In addition, we will
allow for the presence of a cosmological constant $\Lambda$. The
inclusion of this type of matter source was discussed in
\cite{g3}. A previous analysis of the coupling between fermions
and Ashtekar variables can be found in \cite{ferm}.

If we denote the fields $\{\phi,\xi,\bar\eta,A_a\}$ as $\{q^k\}$ and
their respective canonical momenta $\{\pi_{\phi},\rho,\omega, E^a\}$
as $\{p_k\}$, the Lorentzian action can be written as \cite{as91,g3}:
\[
S=\int dt d^3x \big(i\sqrt 2 \sigma^a\dot a_a
+\sum_k   p_k\dot q^k + N {\cal S}+\sum_l u^l\chi_l\big),
\]
where traces over Yang-Mills and $SU(2)$ spinor indices have
been omitted. We have called $\{\chi_l\}$ to the set of
constraints $\{{\cal G}, G,{\cal V}_a\}$. ${\cal G}$ and $G$
denote the Gauss constraints associated with the Ashtekar
connection and the Yang-Mills field, respectively, and ${\cal
V}_a$ is the vector constraint (related to spatial
diffeomorphisms). The set $\{u^l\}$ denotes the Lagrange
multipliers $\{\lambda,\beta,N^a\}$ corresponding to these
constraints, $N^a$ being the shift vector and
$\lambda={^4a}\cdot t$, $\beta=g({^4\!A}\cdot t)$ being the
timelike components of the four-dimensional Ashtekar and
Yang-Mills connections, the latter rescaled by the Yang-Mills
coupling constant $g$ (while the Yang-Mills Gauss constraint has
been rescaled by $g^{-1}$ as compared with that in \cite{as91}).
Finally, ${\cal S}$ is the scalar constraint and $N$ is the
densitized lapse function (with weight equal to $-1$). These
constraints are the generators of the symmetries of the system,
which have their origin in geometrodynamics and Yang-Mills
theory.

In terms of the phase space variables and Lagrange multipliers,
the spacetime line element can be expressed as
\[
ds^2=-\sigma^2 N^2dt^2+h_{ab}(dx^a+N^adt)(dx^b+N^bdt),
\]
where $\sigma=\sqrt{{\rm det}( \sigma^a)}$ and the induced
three-metric $h_{ab}$ is the inverse of
$h^{ab}=-{\rm tr}(\sigma^a\sigma^b)/\sigma^2$.

For the action proposed by Ashtekar, Romano and Tate
\cite{as91,g3}, the explicit form of the constraints can be
found in the appendix. This action and constraints lead to the
Einstein-Cartan theory, which is quartic in the fermionic
variables. Nevertheless, one can attain the Einstein-Dirac
formulation, quadratic in fermionic variables, by simply adding
a fermionic term ${\cal S}_{\rm f}$ to the scalar constraint
\cite{as91}. This term can be obtained from \cite{as91} after
some calculations and has the form
\begin{equation}
{\cal S}_{\rm f}=-\frac{3}{16}  \big( y_A^{\;\;A}  y_B^{\;\;B}+
 y_{AB}  y^{AB} + y_{AB} y^{BA}\big),
\label{sf}
\end{equation}
with
\begin{equation}
 y_{AB}=  \rho_A\xi_B+ \omega_A\bar\eta_B.
\label{yab}
\end{equation}

It is worth noting that the gravitational Gauss constraint
${\cal G}$ does not depend on any external parameter. On the
other hand, the constraints $G$ and ${\cal V}_a$ depend on the
Yang-Mills coupling $g$, and the scalar constraint ${\cal S}$
depends on this constant, the fermion mass $m$, the mass squared
of the scalar field $\mu^2$ and the cosmological constant
$\Lambda$. We will refer to these parameters as coupling
constants and collectively denote them as $\kappa$:
$\kappa\equiv (m,\mu^2,g,\Lambda)$.

The rest of the paper is organized as follows. Section 2 is
devoted to the Wick rotation in general relativity. Constant
scale transformations are studied in section 3. Section 4 deals
with the generalization of the Thiemann transform for gravity
with matter fields. We show that the geometric interpretation
obtained in \cite{me98} for the Thiemann transform in vacuum can
be extended to the case when matter sources are present. We
conclude and summarize in section 5. In the appendix, we give
the detailed expressions of the action and constraints proposed
by Ashtekar, Romano and Tate \cite{as91,g3}.

\section{Wick rotation}

In quantum field theory in flat spacetime, all correlation
functions of physical observables can be obtained by means of
path integrals of the form $\int {\cal D}q e^{iS}$, with ${\cal
D}q$ being a suitable integration measure for the matter fields.
The integrand of this expression is oscillatory and therefore
the path integral formally diverges. To avoid this problem, one
can perform a Wick rotation, i.e. an analytic continuation
$t\rightarrow -it_\euc$ from the Lorentzian time $t$ to the
Euclidean time $t_\euc$. Hence, the spacetime metric acquires
Euclidean signature. In this way, the Lorentzian configurations
are replaced with Euclidean histories and the Lorentzian action
$S$ is mapped to $S(t\rightarrow -it_\euc)= iI(t_\euc)$, where
$I$ is the Euclidean action. It can then be seen that, for
integer-spin fields, $I$ is positive definite and therefore the
Euclidean path integral $\int {\cal D}q e^{-I}$ converges. For
fermions, $I$ is complex, but the convergence of the Euclidean
path integral is also guaranteed provided that one adopts
Berezin integration rules \cite{bere}.

In general relativity, one can similarly relate the Lorentzian
and Euclidean path integrals by means of a Wick rotation. This
gravitational Wick rotation cannot be considered as a rigorously
defined transformation that sends each real Lorentzian metric to
a real Euclidean one. In fact, the analytic continuation in time
of a Lorentzian metric does not generally possess a section in
the complexified spacetime on which the metric is real and has
Euclidean signature \cite{ha84}. Nevertheless, it is still
possible to regard the Wick rotation as a series of substitution
rules that map functions that depend on the real Lorentzian time
to functions that depend on the real Euclidean time coordinate.
Once this point of view is adopted, the gravitational Wick
rotation can be understood as a transformation $R$ that sends,
at least formally, the abstract Lorentzian line element to the
Euclidean one: $R\circ ds^2=ds_\euc^2$, where
\[
ds^2_\euc=\sigma_\euc^2 N_\euc^2dt_\euc^2+
h^\euc_{ab}(dx^a+N_\euc^adt_\euc)(dx^b+N_\euc^bdt_\euc).
\]
In addition, the Lorentzian action $S$ is mapped by this
transformation to
\[
R\circ S=iI.
\]

There are various ways of implementing this Wick rotation. For
instance, we can formally follow the procedure employed in flat
spacetime, namely, to perform the analytic continuation of the
time parameter $t\rightarrow -it_\euc$ \cite{7}. The Lorentzian
line element is then transformed into its Euclidean counterpart
provided that $R\circ N=N_\euc$, $R\circ N^a=iN_\euc^a$ and
\[
R\circ \sigma^a= \sigma^a_\euc.
\]
Lorentzian and Euclidean quantities depend on $t$ and $t_\euc$,
respectively. On the other hand, the action of $R$ on the matter
field variables can be defined by
\[
R\circ q^k=q^k_\euc.
\]
This definition is a natural generalization of the continuation
of the matter fields from the Lorentzian to the Euclidean sector
performed in the flat case.

Alternatively, we can rotate the lapse function to the negative
imaginary axis \cite{g9,g8,hd} and leave the time coordinate and
the shift function unchanged, i.e. $R\circ t=t_\euc$ and
\[
R\circ N=-iN_\euc,
\hspace{5mm}
R\circ N^a=N_\euc^a,
\]
so that the line element is properly transformed under $R$.
With this prescription, $\sigma^a$ and $q^k$ are also left
unchanged by the action of $R$.

Note that the transformation laws for $Ndt$ and $N^adt$ under the
Wick rotation are independent of the specific way in which this
rotation is implemented:
\[
R\circ (N dt)=-iN_\euc dt_\euc,
\hspace{5mm}
R\circ (N^adt)=N^a_\euc dt_\euc.
\]

Finally, we will understand that the action of $R$ on an
integral over time (like, for example, the Lorentzian action) is
given by $R\circ\int dt f(t)\equiv\int R\circ[dt f(t)]$,
regardless of the prescription adopted for the Wick rotation.

If we define the Euclidean phase space action functional $I$ as
\[
 I=\int dt_\euc d^3x \big(\sqrt 2 \sigma_\euc^b\dot a^\euc_b
+\sum_k  p^\euc_k\dot q_\euc^k +N_\euc {\cal S}_\euc
+\sum_lu^l_\euc \chi_l^\euc\big),
\]
we can then deduce the form of the Euclidean matter momenta and
the transformation law for the Ashtekar connection by comparing
the terms in $R\circ S$ and $iI$ that contain time derivatives:
\[
R\circ   p_k=i   p_k^\euc,
\hspace{5mm}
R\circ a_a=a_a^\euc.
\]

In addition, the Lagrange multipliers $u^l$, multiplied by $dt$,
transform according to
\[
R\circ (u^l dt)=u^l_\euc dt_\euc.
\]
Indeed, we have already shown that the product $N^adt$ is not
modified by the Wick rotation. On the other hand, both $\lambda$
and $\beta$ are defined by the scalar product of the vector
$t^a$ and a four-dimensional connection (${^4a}$ and ${^4\!A}$,
respectively). If we then extend the action of the Wick rotation
on the spatial connections $a_a$ and $A_a$ to their
four-dimensional counterparts and take into account that
$t^a\nabla_at=1$, we see that $\lambda dt=(^4 a\cdot t)dt$ and
$\beta dt=g(^4 \! A\cdot t)dt$ are, in fact, unaffected by the
Wick rotation, performed either via a continuation of the time
coordinate or by rotating the lapse function.

We can also see that the Euclidean constraints must be
proportional to the transform of their Lorentzian counterparts
under $R$. The factors that relate both sets of constraints can
be determined in the following way. If we compare the constraint
parts of $R\circ S$ and $iI$ and take into account the
transformation laws for $N dt$ and $u^ldt$, we obtain
\[
R\circ {\cal S}=-  {\cal S}_\euc,
\hspace{5mm}
R\circ  \chi_l=i  \chi_l^\euc,
\]
where ${\cal S}$ and $\chi_l$ depend on the Lorentzian phase space
variables, while ${\cal S}_\euc$ and $\chi_l^\euc$ depend on the
Euclidean ones. Explicitly, these formulae can be written in the
form ${\cal S}(R\circ z)=-{\cal S}_\euc(z_\euc)$ and
$\chi_l(R\circ z)=i\chi_l^\euc(z_\euc)$, $z$ and $z_\euc$
denoting the Lorentzian and Euclidean phase space variables,
respectively.

Notice that, as far as the action, constraints (possibly
including the contribution (\ref{sf}) to the scalar constraint),
and canonical variables are concerned, both ways of implementing
the Wick rotation give equivalent results. Since, in the Hamiltonian
formulation, the system is completely determined by the values
of the phase space variables and Lagrange multipliers on any
surface of constant time, it is nonetheless most convenient to
define the transformation $R$ without making any reference to the
time coordinate. This can be achieved by implementing the Wick
rotation as a continuation of the lapse function to the negative
imaginary axis, thus leaving the time coordinate unaffected.

Let us finally discuss the effect of the Wick rotation on the
extrinsic curvature $K_{ab}$. The Ashtekar connection can be
written as \cite{as91}:
\[
a_a=\Gamma_a-iK_a+C_a,
\]
\begin{equation}
a_a^\euc=\Gamma_a^\euc+K_a^\euc+iC_a^\euc,
\label{eec}
\end{equation}
where (both in the Lorentzian and the Euclidean theories) $K_a=
\sigma^bK_{ab}/(\sqrt 2 \sigma)$, $\Gamma_a$ is the spin
connection compatible with $\sigma^a$ \cite{me98}, and
$C_a$ is the fermionic contribution \cite{as91}
\[
C_a^{\;AB}=\frac{-i}{4\sqrt 2}
(\sigma_{a}^{\;AC} y_C^{\;\;B}+ \sigma_{a}^{\;BC} y_C^{\;\;A}).
\]
In this formula, $\sigma_{a}$ is the inverse of $\sigma^a$ and
$ y_{AB}$ is defined in equation (\ref{yab}). From the transformation
rule for the Ashtekar connection, it is then possible to see that
$R\circ K_a=iK_a^\euc$. Therefore, the extrinsic curvature
transforms under the Wick rotation via
\[
R\circ K_{ab}=iK_{ab}^\euc.
\]
This is precisely the result that one would expect from the
expression for the extrinsic curvature that follows from the
equations of motion:
\begin{equation}
K_{ab}=\frac{1}{2\sigma N}[\dot h_{ab}-2D_{(a}N_{b)}].
\label{kext}
\end{equation}
Here, $D_a$ is the covariant derivative compatible with
$h_{ab}$, $N_b=N^ch_{bc}$, and the parentheses denote
symmetrization of indices.

\section{Constant scale transformations}
\label{conformal}

In this section, we will study the behaviour of the gravitational
and  matter fields under constant scale transformations. With this
aim, we first carry out a dimensional analysis of the various
fields and coupling constants of the theory.

The line element has dimensions of length squared, i.e.
$D(ds^2)=2$. If we describe the spacetime coordinates by means
of dimensionless parameters, the metric tensor inherits the
dimensions of $ds^2$, so that $D(\sigma^a)=2$, $D(N)=-2$, and
$D(N^a)=0$. The spin connection $\Gamma_a$ compatible with $
\sigma^a$ has the same dimension as the derivative operator.
Since the coordinates are dimensionless, so must be $\Gamma_a$.
Using equation (\ref{kext}), it is easy to see that the variable
$K_a= \sigma^bK_{ab}/(\sqrt 2 \sigma)$ is also dimensionless.
Consequently, we obtain $D(a_a)=0$ in vacuum. Actually, one
could have expected this result from the fact that the Ashtekar
connection is invariant under constant scale transformations of
the metric. It is then not difficult to check that the
gravitational action in vacuum has dimensions of area. This
conclusion can also be extended to the full action, for gravity
with matter sources, taking into account the covariance of the
theory and the homogeneity of the action. Thus, we have
$D(S)=2$. Note that this fact naturally selects a parameter
$l_*^2$ with units of area for the path integral (whose weight
would be $\exp(iS/l_*^2)$) and that becomes the expansion
parameter in the semiclassical approximation. This parameter can
be naturally identified with Planck length squared (up to
irrelevant numerical factors).

Since $D(S)=2$, the term $p_k\dot{q}^k$ in the action, which
provides the symplectic structure for the matter fields,
requires that $D(q^k)+D( p_k)=2$. Noting then that the
dimensions of the gravitational variables are not affected by
the presence of matter fields and employing the explicit form of
the constraints given in the appendix, we can easily calculate
the dimensions of all the quantities appearing in the action:
\[
\hspace*{-2mm}
\begin{array}{lllllll}
D(a_a)=0, &&
D(\sigma^a)=2, &  &
D(q^k)= (-1)^{2s_k}s_k,&&
D(p_k)= 2-(-1)^{2s_k}s_k, \\
D(u^l)= 0, &&
D(N)=-2, &&
D(m)=D(g)=-1, &&
D(\mu^2)=D(\Lambda)=-2.
\end{array}
\]
Here, $s_k$ is the spin of the matter field $q^k$, namely,
$s_\phi=0$, $s_\xi=s_{\bar\eta}=1/2$ and $s_\ym=1$. Notice that,
for the gravitational variables, it is the momentum $\sigma^a$
that has the dimension $(-1)^{2s_{\rm g}}s_{\rm g}$ (with
$s_{\rm g}=2)$ instead of the configuration variable $a_a$. This
is because the metric (which can be considered as the physically
relevant gravitational field) is defined in terms of $\sigma^a$
and not in terms of the connection $a_a$. Note also that the
motivation for modifying the Gauss constraint associated with
the Yang-Mills field as well as its Lagrange multiplier by a
factor of $g$ becomes now apparent: the dimension of the
timelike component of the Yang-Mills connection ${^4\!A}\cdot t$
is compensated by introducing the factor $g^{-1}$, so that all
the Lagrange multipliers, except the densitized lapse function,
are dimensionless. On the other hand, the densitized lapse
function must have dimension $-2$ in order to ensure that the
line element has the proper dimensions.

For each complex number $\Omega$, let us now consider the constant
scale transformations defined by means of the operator $C_\Omega$:
\begin{equation}
C_\Omega\circ f=\Omega^{D(f)}f.
\label{csc}
\end{equation}
In particular, for the spacetime metric, the action of $C_\Omega$
amounts to a constant conformal transformation. It is convenient
to split $C_\Omega$ into two pieces, $C_\Omega^{\sc f}$ and
$C_\Omega^{\kappa}$, the former acting only on fields (phase space
variables and Lagrange multipliers) and the latter acting on the
coupling constants $\kappa=(m,\mu^2, g,\Lambda)$. Since the line
element is independent of $\kappa$, we then have
$C_\Omega^{\sc f}\circ ds^2=\Omega^2 ds^2$. In addition, from the
transformation law for the action under $C_\Omega$,
$C_\Omega\circ S=\Omega^2 S$, we obtain
\[
C_\Omega^{\sc f}\circ S(\kappa)=
\Omega^2 (C_\Omega^{\kappa})^{-1}\circ S(\kappa)=
\Omega^2 S[\kappa(\Omega)],
\]
where $\kappa(\Omega)\equiv (m\Omega,\mu^2\Omega^2,
g\Omega,\Lambda\Omega^2)$. Taking then into account the
dimensions of the Lagrange multipliers, we conclude that, acting
on the constraints, $C_\Omega^{\sc f}$ is
\[
C_\Omega^{\sc f}\circ \chi_l(g)
=\Omega^2 \chi_l(g\Omega),
\hspace{5mm}
C_\Omega^{\sc f}\circ { {\cal S}}(\kappa)
=\Omega^4 { {\cal S}}[\kappa(\Omega)].
\]
Therefore, we see that, in general, $C_\Omega^{\sc f}$ transforms
constraints with physical real couplings $\kappa$ into
constraints with complex couplings $\kappa(\Omega)$.

\section{Thiemann transform}
\label{thiemann}

The Thiemann transform is an automorphism on the algebra of
functions on phase space that preserves the symplectic
structure. As such, it can be implemented as a transformation
generated via Poisson brackets, as discussed in
\cite{th96,as96,me98,me97}. None of the transformations
considered so far, namely, the Wick rotation $R$ and the
constant rescaling of the fields $C^{\sc f}_\Omega$, preserves
the symplectic structure. Furthermore, they not only affect the
phase space variables but also the Lagrange multipliers.
However, if we set $\Omega=e^{i\pi/4}$ and denote the
corresponding operator $C^{\sc f}_\Omega$ by $C^{\sc f}$, it
turns out that the composition $T\equiv C^{\sc f}\circ R^{-1}$
commutes with the Poisson bracket operation. Indeed, the
explicit transformation rules for the Euclidean phase space
variables under $T$ are
\[
\hspace*{-1.5mm}
\begin{array}{lll}
T\circ a_a^\euc=a_a, &&
T\circ q^k_\euc= q^k e^{i \pi (-1)^{2s_k}s_k/4},
\\
T\circ\sigma_\euc^a=i\sigma^a, &&
T\circ p_k^\euc= p_k e^{-i \pi (-1)^{2s_k}s_k/4}.
\end{array}
\]

Using then that the replacement of the Euclidean Ashtekar
connection $a_a^\euc$ with the variable
$x_a^\euc=a_a^\euc-\Gamma_a^\euc$ (and the change from $ia_a$ to
$x_a=i(a_a-\Gamma_a)$ in the Lorentzian theory) amounts to a
canonical transformation \cite{as91} and recalling that
$\Gamma_a^\euc$ is invariant under constant rescalings of
$\sigma^a_\euc$, it is not difficult to see that, on phase
space, the transformation $T$ is generated via Poisson brackets
by
\begin{equation}
{\cal T}=\frac{\pi}{4}\sum_\alpha (-1)^{2s_\alpha}s_\alpha
\int d^3x  p_\alpha^\euc q^\alpha_\euc,
\label{gent}
\end{equation}
where $\{q^\alpha_\euc\}$ denotes the set of variables
$\{-\sqrt 2 \sigma^a_\euc,q^k_\euc\}$, their canonically conjugate
variables $\{x_a^\euc, p_k^\euc\}$ are denoted as
$\{p_\alpha^\euc\}$, and $s_\alpha$ is the spin of $q^\alpha_\euc$.
In the above expression for ${\cal T}$, traces over $SU(2)$ and
Yang-Mills indices are implicitly assumed. On functions of the
Euclidean phase space variables, the action of $T$ is then given by
\begin{equation}
T\circ f(q^\alpha_\euc, p_\alpha^\euc)= {\cal I}\circ\sum_n
\frac{i^n}{n!}\{f(q^\alpha_\euc, p_\alpha^\euc), {\cal T}\}_{(n)}.
\label{tfz}
\end{equation}
Here, $\{\cdot, {\cal T}\}_{(n)}$ is the $n$th application of
the Poisson bracket with ${\cal T}$, and ${\cal I}$ is an
isomorphism between (the algebras of functions on) the Euclidean
and Lorentzian phase spaces which, in practical terms, just
removes the index $\euc$ from the canonical variables
$(q^\alpha_\euc, p_\alpha^\euc)$, i.e. ${\cal I}\circ
(q^\alpha_\euc, p_\alpha^\euc)=(q^\alpha,p_\alpha)$.

In addition, if one adheres to the prescription of performing the
Wick rotation by means of a continuation of the lapse function, it
is straightforward to check that $T=C^{\sc f}\circ R^{-1}$
transforms the Euclidean Lagrange multipliers into the Lorentzian
ones
\[
T\circ N^\euc=N,
\hspace{5mm}
T\circ u^l_\euc=u^l.
\]
Note that the definition of $T$ provided by equation (\ref{tfz})
can then be extended to the Lagrange multipliers, since they can
be considered as functions that are independent of the phase
space variables and, consequently, have vanishing Poisson
brackets with any other function on phase space like, for
example, the generator ${\cal T}$.

On the other hand, the constraints transform under $T$ according
to
\[
T\circ {\cal S}_\euc(\kappa)
={\cal S}(\kappa'),
\hspace{5mm}
T\circ\chi_l^\euc(\kappa)=
\chi_l(\kappa'),
\]
where $\kappa'\equiv\kappa(\Omega=e^{i\pi/4})$. Provided that,
similar to what happened with the Wick rotation, one understands the
transformation under $T$ of an integral over time as $T\circ \int
dt f(t)=\int dt T\circ f(t)$, it follows that $T$ maps the Euclidean
to the Lorentzian action, but with different parameters, namely,
\[
T\circ I(\kappa)=S(\kappa').
\]
It should be stressed that, in order to arrive at the above
transformation rules for the constraints and the action, we have
assumed that $T$ does not directly affect the coupling constants
(i.e. $T\circ\kappa=\kappa$). Finally, one can readily check
that the line element acquires a factor of $i$ under the action
of $T$:
\[
T\circ ds^2_\euc=ids^2,
\]
that is, apart from mapping the Euclidean to the Lorentz\-ian
line element, $T$ induces a constant complex conformal
transformation of the spacetime metric.

In view of these transformation laws, we conclude that $T=C^{\sc
f}\circ R^{-1}$ provides in fact the Thiemann transform, since it
is a map on the algebra of functions on phase space that
preserves the Poisson bracket structure and sends the Euclidean
constraints to their Lorentzian counterparts (although with
complexified coupling constants $\kappa'$).

It is worth remarking that the Thiemann transform attained in this
way acts as the identity operator on all the Lagrange multipliers.
If we had instead implemented the Wick rotation by means of the
continuation in time $t\rightarrow -it_\euc$ and had denoted such a
rotation as $\bar R$, to differentiate it from that obtained via a
continuation of the lapse, then the transform
$\bar T=C^{\sc f}\circ\bar R^{-1}$ would act on the Lagrange
multipliers as
\[
\bar T\circ N_\euc=-iN,
\hspace{5mm}
\bar T\circ u^l_\euc=-iu^l.
\]
However, the action of $\bar T$ would actually coincide with that
of the Thiemann transform $T$ on all the Euclidean phase space
variables. As a consequence, both transforms could be considered
equivalent modulo gauge, in the sense that their effects on any
dynamical trajectory on the Euclidean constraint surface would
only differ by a complex gauge transformation.

A different way of obtaining the Thiemann transform (\ref{tfz}),
which generalizes a proposal put forward by Ashtekar
\cite{as96}, is the following. We make the ansatz
\[
{\cal T}=\frac{\pi}{4}\sum_\alpha b_\alpha
\int d^3x  p_\alpha^\euc q^\alpha_\euc
\]
for the generator of the Thiemann transform $T$ and set $b_{\rm
g}=2$ for the gravitational field, so that one recovers the
definitions of $T$ in vacuum \cite{th96}. Furthermore, we also
require that the remaining constants $b_k$ have the smallest
absolute value such that $T$ maps the Euclidean constraints to the
Lorentzian ones, up to a possible change in the coupling constants
$\kappa$. By considering the transform of the scalar constraint, one
then concludes that the unknown constants $b_\alpha$ must take the
values $(-1)^{2s_\alpha}s_\alpha$, so that they are determined by
the spin of the fields $q^\alpha_\euc$. A comparison of our ansatz
for ${\cal T}$ with equation (\ref{gent}), shows that the transform
derived in this way is precisely the Thiemann transform
$T=C^{\sc f}\circ R^{-1}$ discussed above, as we wanted to show.

Ashtekar had already noted that the generator of the Thiemann
transform had the form
\[
\frac{\pi}{4}\sum_\alpha s_\alpha\int d^3x p_\alpha^\euc
q^\alpha_\euc
\]
and that this ``might well be a reflection of a deeper structure
underlying the generalized Wick transform'' \cite{as96}. We have
seen that the spin dependent factor is
$(-1)^{2s_\alpha}s_\alpha$ instead of $s_\alpha$, although a
canonical transformation $(q,p)\rightarrow(-p,q)$ applied to the
fermionic variables relates both expressions. The underlying
structure that Ashtekar refers to is actually present. Indeed,
$T$ preserves Poisson brackets and, therefore, it is sufficient
to know the action of $T$ on the variables $q^\alpha_\euc$ in
order to determine $T$ completely. Moreover, since the Thiemann
transform is a composition of an inverse Wick rotation $R^{-1}$
with the constant scale transformation of the fields $C^{\sc
f}$, and the inverse Wick rotation acts on the variables
$q^\alpha_\euc$ as the trivial isomorphism ${\cal I}$ introduced
in equation (\ref{tfz}), we just need to know the transformation
rules for the variables $q^\alpha_\euc$ under $C^{\sc f}$. In
other words, the Thiemann transform is completely determined by
the action of constant scale transformations on the variables
$q^\alpha_\euc$, in which the dimensions of the fields
$q^\alpha_\euc$ play a central role. Finally, the covariance of
the theory and the fact that the line element has dimension 2
relate the dimension and spin of the different fields.
Summarizing, the requirement that the Thiemann transform
preserves Poisson brackets, together with the covariance of the
theory and the dimensional character of the spacetime line
element, are the key structural features that underlie the
Thiemann transform.

In \cite{as96}, Ashtekar proposed an extension of the Thiemann
transform, defined for vacuum general relativity, to gravity
with a cosmological constant, coupled with a massive scalar
field and an Abelian Yang-Mills field (this Abelian case
corresponds to the limit of vanishing Yang-Mills constant $g$).
Fermions were not included. Such an extension $T_{\sc a}$ was
determined by the requirement that it maps the Euclidean action
functional to the Lorentzian one with identical values of the
coupling constants $\kappa$. In particular, this implies that
$T_{\sc a}$ sends Euclidean to Lorentzian constraints via
\[
T_{\sc a}\circ {\cal S}_\euc(\kappa)={\cal S}(\kappa),
\hspace{5mm}
T_{\sc a}\circ\chi_l^\euc(\kappa)=\chi_l(\kappa).
\]
The transformation laws obtained by Ashtekar for phase space
variables coincide with those found here for $T$ whereas, for
coupling constants and Lagrange multipliers, they have the form
\[
T_{\sc a}\circ\mu^2=-i\mu^2, \hspace*{5mm}
T_{\sc a}\circ\Lambda=-i\Lambda,
\]
\[
T_{\sc a}\circ N_\euc=-N, \hspace*{5mm}
T_{\sc a}\circ N^a_\euc=N^a,\hspace*{5mm}
T_{\sc a}\circ({^4\!A}\cdot t)_\euc=e^{i\pi/4}({^4\!A}\cdot t).
\]

The difference in the sign of the transformation laws for the
densitized lapse function under $T$ and $T_{\sc a}$ simply comes
from a choice of a wrong relative sign between the Euclidean and
Lorentzian scalar constraints as compared with the standard
conventions. With the usual choice of sign, one has
\[
T_{\sc a}\circ N_\euc=N.
\]
On the other hand, the above transformation law for
${^4\!A}\cdot t$ can actually be recovered in terms of the
transform $T$ by noting that ${^4\!A}\cdot t=g^{-1}\beta (g)$ (a
relation that continues to be valid in the limit of vanishing
$g$), taking into account that $T\circ\beta_\euc(g)=\beta(g')$,
and recalling that $T$ does not directly affect the coupling
constants. Finally, notice that both $T$ and $T_{\sc a}$ leave
the shift vector invariant. Therefore, it turns out that, for
the matter content considered in \cite{as96}, the only
difference between $T$ and $T_{\sc a}$ is that the latter
affects the coupling constants $\kappa$ while the former does
not. In fact, it is easy to see that
\[
T_{\sc a}=C^\kappa\circ T,
\]
where $C^\kappa$ is the constant scale transformation with
factor $e^{i\pi/4}$ restricted to act only on $\kappa$. Thus,
while $T=C^{\sc f}\circ R^{-1}$ is a composition of an inverse
Wick rotation and a constant scale transformation of the fields,
$T_{\sc a}$ is equivalent to the composite of an inverse Wick
rotation and a proper rescaling that affects all dimensional
quantities, including the coupling constants, i.e.
\[
T_{\sc a}=C\circ R^{-1},
\]
with $C$ being the transformation $C_{\Omega}$, defined in equation
(\ref{csc}), evaluated at $\Omega=e^{i\pi/4}$.

\section{Conclusions and further comments}

The Thiemann transform was originally defined in vacuum general
relativity as an automorphism on the algebra of functions on
phase space that preserves the Poisson bracket structure and
maps the constraints of Euclidean gravity to their Lorentzian
counterparts. This transform was extended by Ashtekar to the
case in which a scalar and a Maxwell field were present. In this
work, we have further extended the Thiemann transform to the
Ashtekar formulation for general relativity in the presence of
general matter sources, namely, spin-1/2 fermions, a
(non-Abelian) Yang-Mills field and a scalar field. We have
proved that the action of the Thiemann transform on functions of
the phase space variables is in fact equivalent to the result of
an inverse Wick rotation and a constant scale transformation.
Moreover, this equivalence also holds on functions that depend
 on the Lagrange multipliers as well (including the lapse
function and the shift vector), provided that one performs the
Wick rotation by means of an analytic continuation of the lapse
and scales up all gravitational and matter fields, i.e. both the
phase space variables and the Lagrange multipliers. For the
spacetime metric, this complex rescaling amounts to a constant
conformal transformation.

These results endow the Thiemann transform with a geometric
interpretation. In addition, they provide, together with arguments
based on general covariance and the dimensions of the line element,
a simple explanation of the fact that the generator of the Thiemann
transform is essentially determined by the spin of the different
fields that are present in the theory \cite{as96}.

Since the Thiemann transform $T$ relates the Eu\-clid\-ean and
Lorentzian sectors of general relativity, one would expect that,
were it possible to define a quantum analog of $T$, the
Lorentzian quantum states could be obtained by transforming the
solutions of the Euclidean quantum constraints \cite{as96}. In
this way, the Thiemann transform could be employed in the
quantum theory to circumvent the problem of imposing complicated
reality conditions on the Ashtekar variables for gravity,
because one can actually take these variables as real in the
Euclidean sector (for an alternative procedure that avoids the
use of a complex-valued Ashtekar connection in Lorentzian
quantum gravity, see \cite{th98}). More precisely, if one uses
the Pauli matrices $\tau_{\;A}^{i\;\;B}$ (with $i=1,2,3$) to
express the Euclidean Ashtekar variables in the form \cite{as91}
\[
\sigma^a_{\euc}=-\frac{i}{\sqrt{2}}\sigma^a_i\tau^i,
\hspace*{5mm} a_a^{\euc}=-\frac{i}{2}a_a^i\tau^i,
\]
it turns out that the densitized triad $\sigma^a_i$ and the
connection $a_a^i$ can be chosen to be real. In the absence of
fermions this is a consequence of the fact that the inclusion of
matter sources does not modify the reality conditions for the
gravitational variables, so that, as in vacuum Euclidean general
relativity, one can restrict $\sigma^a_i$ and $a_a^i$ to be real
\cite{as96}. In the presence of fermions, one can still work
with real Euclidean triads $\sigma^a_i$. However, the reality
conditions proposed in \cite{as91,g3} for the fermionic fields,
together with relation (\ref{eec}) and the reality of the
Euclidean extrinsic curvature, can be shown to imply that the
connection $a_a^i$ has then an imaginary part equal to
$-i\sigma_a^i{\rm tr}(y)/4$, where $\sigma_a^i$ is the inverse
of the densitized triad $\sigma^a_i$ and $y_{AB}$ is the
function defined in equation (\ref{yab}). Nevertheless, one can
easily recover a real connection by simply replacing $a_a^i$
with its real part, namely, with $a_a^i-\sigma_a^i{\rm
tr}(y)/4$. Indeed, it is not difficult to check that this
replacement just amounts to a canonical transformation on the
Euclidean phase space when it is accompanied with the following
change of fermionic variables
\[
(\xi_{\euc},\rho_{\euc},\bar{\eta}_{\euc},\omega_{\euc})
\rightarrow
(\sigma^{1/2}\xi_{\euc},\sigma^{-1/2}\rho_{\euc}
,\sigma^{1/2}\bar{\eta}_{\euc},\sigma^{-1/2}
\omega_{\euc}).\]

It is remarkable that the Thiemann transform may provide a way
to extract the Lorentzian quantum physics from Euclidean general
relativity. In this sense, we note that the Wick rotation for
gravity is rather a formal technique for passing from the
Lorentzian to the Euclidean path-integral approach. On the other
hand, it is well known that quantum gravity is
non-renormalizable, so that a change of scale cannot be simply
absorbed by a redefinition of the coupling constants. However,
we have seen that the combination of an inverse Wick rotation
with a change of scale by a fixed factor of $e^{i\pi/4}$ turns
out to provide a rigorously defined transformation, which is
just the Thiemann transform. Thus, if a quantum analogue of this
transform exists, it should describe the behaviour of the
quantum states under the simultaneous implementation of a Wick
rotation and a fixed constant scale transformation of all
fields, even if these two kinds of transformations are
separately ill-defined in the quantum theory.

A line of research in which the Thiemann transform might find
applications is the study of gravitational thermodynamics. Since
statistical mechanics is naturally formulated in the Euclidean
sector, the Thiemann transform might well be a useful tool to
obtain the thermodynamical properties of Lorentzian gravity from
an Euclidean formalism. One reason why classical solutions in
gravity have intrinsic entropy is that the gravitational action
is not scale invariant, but behaves instead like an area under a
change of scale in the solutions \cite{7}. As we have seen, this
fact plays a key role in the existence of the Thiemann transform
and its geometric interpretation. In addition, it is worth
noticing that the thermodynamical properties of a gravitational
system in vacuum are determined in the semiclassical
approximation by the action of the solutions. This action comes
entirely from a surface contribution that is given by the
integral of the densitized trace of the extrinsic curvature
\cite{7}. Quite remarkably, this surface term has formally the
same expression as the generator of the Thiemann transform in
vacuum general relativity \cite{th96}.

\section*{Acknowledgments}

The authors are grateful to P. F. Gonz\'{a}lez D\'{\i}az for helpful
conversations. This work was supported by funds provided by the
DGICYT Project No. PB94--0107.

\appendix

\section*{Appendix}

Let us give the explicit expression of the self-dual action
employed in \cite{as91,g3} to discuss the introduction of matter
sources in the Ashtekar formalism for gravity. In the
Hamiltonian formulation, this action can be written in the form
\begin{eqnarray}
 S^\epsilon=
\int dt d^3x &&\hspace*{-6mm}\big\{
\epsilon^{-1}\sqrt 2 {\rm tr} (\sigma^a\dot a_a)
+\pi_{\phi}\dot\phi+{\rm tr}( E^a\dot A_a)
+\rho_A\dot\xi^A+\omega_A\dot{\bar\eta}^A
\nonumber\\
&&\hspace*{-6mm}
+ g{\rm tr}[({^4\!A}\cdot t)G^\epsilon]
+{\rm tr }[({^4a}\cdot t){\cal G}^\epsilon]
 +N^a{\cal V}^\epsilon_a+N{\cal S}^\epsilon\big\},
\nonumber
\end{eqnarray}
where traces have been displayed explicitly. In this formula,
$\epsilon=-i$ for the Lorentzian theory, while the Euclidean action
is obtained by setting $\epsilon=1$. The Gauss constraint associated
with the Yang-Mills connection takes the same expression in the
Lorentzian and Euclidean sectors, namely,
\[
G^\epsilon=g^{-1}D_a E^a.
\]
The Gauss constraint associated with the Ashtekar connection and
the vector constraint can be written as
\begin{eqnarray}
{\cal G}^\epsilon_{AB}=
&&\hspace*{-6mm}
\frac{1}{\epsilon}\sqrt 2 {\cal D}_a \sigma^a_{AB}
-\rho_{(A}\xi_{B)} -\omega_{(A}\bar\eta_{B)},
\nonumber\\
{\cal V}^\epsilon_a=
&&\hspace*{-6mm}
-\frac{1}{\epsilon}\sqrt 2{\rm tr}(\sigma^b F_{ab})
-\rho_A {\cal D}_a \xi^A  -\omega_A {\cal D}_a\bar\eta^A
-\pi_{\phi}\partial_a\phi -\frac{1}{2} {\rm tr} ( E^b B_{ab}).
\nonumber
\end{eqnarray}
Finally, the scalar constraint is given by
 \begin{eqnarray}
 \epsilon^2{\cal S}^\epsilon= &&\hspace*{-6mm} {\rm tr}
 (\sigma^a\sigma^b F_{ab}) +\frac{1}{8\sigma^2}{\rm tr}
 (\sigma^a\sigma^c ) {\rm tr}(\sigma^b\sigma^d ) {\rm tr}
 (\epsilon^2 E_{ab}E_{cd}-B_{ab}B_{ cd})
 \nonumber\\
 &&\hspace*{-6mm} -i
 m(\sigma^2\xi^A\bar\eta_A+\epsilon^2\rho^A\omega_A)
 +\epsilon\sqrt 2  \sigma_{\;A}^{a\;\;B}
 (\rho_B{\cal D}_a\xi^A+\omega_B{\cal D}_a\bar\eta^A)
 \nonumber\\
 &&\hspace*{-6mm} +\epsilon^2\frac{\pi_{\phi}^2}{16\pi}
 -4\pi\sigma^2\mu^2\phi^2 +4\pi{\rm tr}(\sigma^a\sigma^b)
 \partial_a\phi\partial_b\phi -\sigma^2\Lambda.
 \nonumber
 \end{eqnarray}

In these expressions, $D_a$ is the derivative operator associated
with the Yang-Mills connection $A_a$, and $B_{ab}$ is twice its
curvature:
\begin{eqnarray}
D_a E^a=
&&\hspace*{-6mm}
\partial_a E^a+g [A_a,E^a],
\nonumber\\
B_{ab}=
&&\hspace*{-6mm}
2(\partial_a A_b-\partial_b A_a+g[A_a,A_b]).
\nonumber
\end{eqnarray}
In addition, ${\cal D}_a $ is the derivative operator associated
with the Ashtekar connection $a_a$, and $F_{ab}$ is its
curvature, i.e.
\settowidth{\inda}{$^{aa}$}
\settowidth{\indb}{$^{aaa}$}
\begin{eqnarray}
{\cal D}_a \xi^A=
&&\hspace*{-6mm}
\partial_a \xi^A
-a_{a\;B}^{\;A}\xi^B,
\nonumber\\
F_{abA}^{\hspace{\indb}B}=
&&\hspace*{-6mm}
\partial_a a_{bA}^{\hspace{\inda}B}
-\partial_b a_{aA}^{\hspace{\inda}B}
+a_{aA}^{\hspace{\inda}C}a_{bC}^{\hspace{\inda}B}
-a_{bA}^{\hspace{\inda}C}a_{aC}^{\hspace{\inda}B}.
\nonumber
\end{eqnarray}
Finally, $\sigma=\sqrt{{\rm det}(\sigma^a)}$ and
\[
E_{ab}=\eta_{abc} E^c,
\]
where $\eta_{abc}$ is the c-number Levi-Civita form-density.

\end{document}